\documentclass[12pt]{article}

\newfont{\Bbb}{msbm10 scaled 1200}
\newcommand{\mathbb}[1]{\mbox{\Bbb #1}}

\def\i{{\mathrm i}}
\def\e{{\mathrm e}}

\def\Z{{\mathbb Z}}

\def\tr{\mathop{\mathrm{tr}}}

\def\sgn{\mathop{\mathrm{sgn}}}
\def\mod{\mathop{\mathrm{mod}}}

\def\be{\begin{equation}}
\def\ee{\end{equation}}
\def\bea{\begin{eqnarray}}
\def\eea{\end{eqnarray}}
\def\ba{\begin{array}}
\def\ea{\end{array}}

\def\Si{{{\displaystyle\sum\mkern-20mu\int\,}}}
\def\si{\mathop{\Si}}
\def\la{\langle}
\def\ra{\rangle}

\newcommand{\bpartial}{\mbox{\boldmath $\partial$}}
\newcommand{\balpha}{\mbox{\boldmath $\alpha$}}
\newcommand{\bgamma}{\mbox{\boldmath $\gamma$}}
\newcommand{\bLambda}{\mbox{\boldmath $\Lambda$}}
\newcommand{\bSigma}{\mbox{\boldmath $\Sigma$}}
\newcommand{\bsigma}{\mbox{\boldmath $\sigma$}}
\begin{document}

\title{\bfseries Induced quantum numbers of \\
a magnetic monopole at finite temperature}
\author{Yu.A.~Sitenko$^{1,2}$, A.V.~Solovyov$^{1,3}$, and
N.D.~Vlasii$^{1,2}$}
\date{}
\maketitle

\begin{center}
\slshape $^{1}$Bogolyubov Institute for Theoretical Physics, \\
National Academy of Sciences, 03680, Kyiv 143, Ukraine \\
$^{2}$Physics Department, National Taras Shevchenko University of
Kyiv, \\
03127, Kyiv 127, Ukraine \\
$^{3}$Department of Physics, Princeton University, \\
Princeton, NJ 08544, USA
\end{center}

\bigskip

\begin{abstract}

A Dirac electron field is quantized in the background of a Dirac
magnetic monopole, and the phenomenon of induced quantum numbers in
this system is analyzed. We show that, in addition to electric
charge, also squares of orbital angular momentum, spin, and total
angular momentum are induced. The functional dependence of these
quantities on the temperature and the CP-violating vacuum angle is
determined. Thermal quadratic fluctuations of charge and squared
total angular momentum, as well as the correlation between them and
their correlations with squared orbital angular momentum and squared
spin, are examined. We find the conditions when charge and squared
total angular momentum at zero temperature are sharp quantum
observables rather than mere quantum averages.
\newline PACS numbers: 14.80.Hv, 11.30.Er, 11.10.Wx
\end{abstract}

\section{Introduction}

The interaction of quantized Dirac fields with classical background
fields of nontrivial topology can give rise to quantum states with
rather unusual eigenvalues~\cite{Jac1,Jac2,Gol1,Nie1,Par,Far}. In
particular, the ground state of a Dirac electron in the background
of a pointlike magnetic monopole acquires nonzero electric charge,
and this results in the monopole becoming a CP symmetry violating
dyon~\cite{Wit,Gro,Yam}. The effect persists when thermal
fluctuations of the quantized Dirac electron field are taken into
account, and this yields temperature dependence of the induced
charge~\cite{Cor,Dun}.

The aim of the present paper is to show that, in addition to charge,
also other quantum numbers are induced in the magnetic monopole
background both at zero and nonzero temperatures. We find
relationships between all these quantum numbers and discuss, which
of them become sharp quantum observables rather than quantum
averages and also when this happens. At nonzero temperature all
quantum numbers are not sharp observables, but, instead, are thermal
averages; and, appropriately, the thermal quadratic fluctuations are
nonvanishing. If a quadratic fluctuation vanishes at zero
temperature, then a corresponding quantum number at zero temperature
becomes a sharp observable. We find out, in particular, that induced
charge and squared total angular momentum at zero temperature are
sharp observables for almost all values of the vacuum angle with the
exception of the one corresponding to zero energy of the bound state
in the one-particle electron spectrum.

A configuration of a pointlike monopole with magnetic charge $g$ at
the origin is given by the field strength in the form \be {\bf
B}({\bf r})= g\frac{\bf r}{r^3},\quad {\bpartial}\cdot{\bf B}({\bf
r})= 4\pi g\delta^3({\bf r}). \ee Although in the space outside the
monopole (i.e. the punctured space that is characterized by the
nontrivial second homotopy group, $\pi_2=\Z$, where $\mathbb{Z}$ is
the set of integer numbers) the magnetic field satisfies the usual
sourceless equation, due to some cohomological obstacles the gauge
vector potential can be defined only \emph{locally}. When attempting
to extend the local potential to a global single-valued one, a
singularity on a halfline or otherwise (so-called Dirac string) is
inevitably encountered \cite{Di}. Namely the condition of
unobservability of the Dirac string yields quantization of monopole
charge $g$.

It should be noted that the Dirac quantization was obtained by
Jackiw~\cite{Jac} in a different way, as a consequence of associativity of spatial
translations in quantum mechanics. Thus, his derivation is
completely gauge-invariant, dispensing with reference to a vector
potential; moreover, it demonstrates in addition that magnetic
monopoles have to be structureless point objects.

Following Wu and Yang~\cite{WY1}, one can introduce the patched gauge vector potential
which is free of singularities. The punctured space is divided into
two overlapping regions,
$R_a:0<\vartheta<\frac{\pi}2+\delta$, and $R_b:\frac\pi
2-\delta<\vartheta<\pi$ ($0\leq\vartheta\leq\pi$ stands for the
azimuthal angle in spherical coordinates,
$x=r\sin\vartheta\cos\phi$, $y=r\sin\vartheta\sin\phi$,
$z=r\cos\vartheta$, and $0<\delta<\frac\pi 2$), and the vector
potential is defined for each of the regions: \be {\left\{
\begin{array}{lr}
{\bf A}({\bf r})\cdot d{\bf r}=g(1-\cos\vartheta)d\phi, & {\bf r}\in R_a,\\
{\bf A}({\bf r})\cdot d{\bf r}=-g(1+\cos\vartheta)d\phi, & {\bf r}\in
R_b,
\end{array}
\right.} \ee then $\bpartial\times{\bf A}={\bf B}$, where $\bf B$ is
given by Eq.(1). In the overlap
$R_{ab}:\frac\pi2-\delta<\vartheta<\frac{\pi}2+\delta$, the two
potentials are related by gauge transformation \be {\bf A}|_a={\bf
A}|_b + \frac{\i}{e}S_{ab}{\bpartial}S_{ab}^{-1}, \ee with \be
\label{GT} S_{ab}=\e^{2\i eg\phi}, \ee $e$ is the electron charge.
Therefore, the vector potential serves as a connection on a
nontrivial $U(1)$ bundle, and the electron wave function  is a
section of this bundle, i.e. wave function $\Psi({\bf r},\,t)$ is
two-valued with its values in the overlap $R_{ab}$ related by gauge
transformation \be\Psi|_a=S_{ab}\Psi|_b.\ee Generating function
$S_{ab}$ (4) is existing (i.e. single-valued) only when \be
\label{Dirac condition} eg=\frac12n,\qquad n\in \Z, \ee which is the
celebrated Dirac quantization condition~\cite{Di} that has already
attained its 75-year anniversary.

Schwinger~\cite{Schw} and Zwanziger~\cite{Zw} generalized condition
(6) to allow for the possibility of particles that carry both
electric and magnetic charges (dyons). A quantum-mechanical theory
can have two particles of electric and magnetic charges $(Q_1,g_1)$
and $(Q_2,g_2)$ only if \be \label{DSZ} Q_1 g_2 - Q_2 g_1 =\frac12
n. \ee Since the electron has no magnetic charge, the quantization
condition says nothing about the electric charge of a dyon. The
quantization condition does say something about the difference
between the electric charges of two dyons. Given, for instance, two
dyons of minimally allowed magnetic charge $g=(2e)^{-1}$ and of
electric charges $Q_1$ and $Q_2$, one gets \be Q_1-Q_2=ne,\ee but
there is no restriction on $Q_1$ and $Q_2$ \emph{separately}. If,
however, the Dirac-Schwinger-Zwanziger quantization condition,
Eq.(7), is supplemented by CP conservation, then the allowed values of
the electric charge of a dyon are quantized and restricted to be
either integer or half-integer in units of $e$. This is due to the
fact that the electric charge is odd and the magnetic charge is even
under CP.

The effect of CP violation was
analyzed by Witten~\cite{Wit} in the framework of a spontaneously
broken gauge theory at nonzero vacuum angle $\Theta$. By introducing
the $\Theta$-term which causes CP violation in the lagrangian, \be
\Delta
L=\Theta\frac{e^2}{(8\pi)^2}\varepsilon^{\mu\nu\mu'\nu'}F_{\mu\nu}
F_{\mu'\nu'}
\ee
(here $F_{\mu\nu}$ is the gauge field strength
and $\varepsilon^{\mu\nu\mu'\nu'}$ is the totally antisymmetric tensor),
he got the expression for the dyon charge \be \label{Witten formula}
Q=ne-\frac{e\Theta }{2\pi}. \ee

On the other hand, the naive Dirac hamiltonian for the electron in
the background of a pointlike magnetic monopole appears to be
non-self-adjoint, and an extra boundary condition at the location of
the monopole is required for the lowest partial wave in order to
implement a self-adjoint extension. The boundary condition depends
on self-adjoint extension parameter $\Theta$ which violates CP
invariance. By quantizing the electron field in the monopole
background and considering the appropriate vacuum polarization
effects, Grossman~\cite{Gro} and Yamagishi~\cite{Yam} got the
expression for the induced vacuum charge \be
Q=-2e|eg|\,\frac1\pi\arctan\left(\tan\frac\Theta2\right), \ee which
in the case of the minimal monopole strength, $g=\pm(2e)^{-1}$,
agrees with Eq.(10). Thus, in this approach the self-adjoint
extension parameter plays the role of the vacuum angle in Witten's
approach and the monopole becomes a dyon owing to the vacuum
polarization effects.

In the present paper we proceed further and find
other quantum numbers of the fermionic vacuum and of the
fermionic system in thermal bath in the monopole background. Similar
problems were considered for planar fermionic systems in the background of a
pointlike magnetic vortex in Refs.\cite{Sit1,Sit2}.

\section{Operators of physical observables and their vacuum and thermal expectation
values}

For a given classical static background field configuration, the
second-quantized fermion field operator $\Psi$ can be expanded in a
complete set of eigenstates of the Dirac (one-particle) hamiltonian
$H$, see, e.g., Ref.\cite{Pes}, \be \label{quantized field}
\Psi({\bf r},t)=\si_{(E_{\lambda}>0)}e^{-\i E_{\lambda}t}\langle
{\bf r}|\lambda\rangle a_{\lambda}+ \si_{(E_{\lambda}<0)}e^{-\i
E_{\lambda}t}\langle {\bf r}|\lambda\rangle b^\dagger_{\lambda} \,,
\ee where \be \label{Dirac_equation} H\langle{\bf r}|\lambda\rangle
=E_{\lambda}\langle {\bf r}|\lambda\rangle, \ee is the stationary
Dirac equation with eigenvalues of $H$ denoted by $E_\lambda$,
$\lambda$ stands for the set of parameters (quantum numbers)
specifying a one-particle state, and symbol $\si$ means the
summation over discrete and the integration (with an appropriate
measure) over continuous values of $\lambda$; $a^\dagger_{\lambda}$
and $a_{\lambda}\;(b_{\lambda}^\dagger \,\mathrm{and}\,b_{\lambda})$
are the fermion (antifermion) creation and destruction operators
obeying anticommutation relations \be
[a_{\lambda},a^\dagger_{\lambda'}]_+
=[b_{\lambda},b_{\lambda'}^\dagger]_+
=\langle\lambda|\lambda'\rangle, \ee and ground state $|{\rm
vac}\rangle$ of the second-quantized theory is defined as \be
a_\lambda|{\rm vac}\rangle=b_\lambda|{\rm vac}\rangle=0. \ee

In the second-quantized theory, the operator of a dynamical variable
(physical observable) is given by the integrated commutator, \be
\hat{O}_\Upsilon=\frac12\int d^3r\,{\rm tr}[\Psi^+({\bf
r},t),\,\Upsilon\Psi({\bf r},t)]_{-}, \ee where $\Upsilon$ is the
appropriate one-particle operator in the first-quantized theory, and
$tr$ denotes the trace over spinor indices; in particular,
$\hat{O}_H$ is the operator of energy, and $\hat{O}_I$ is the
operator of fermion number, where $I$ is the unity matrix in spinor
indices. The vacuum expectation value of the observable
corresponding to Eq.(16) can be presented as \be \langle{\rm
vac}|\hat{O}_\Upsilon|{\rm vac}\rangle=-\frac12{\rm Tr}\Upsilon{\rm
sgn}(H), \ee where $ {\sgn}(u)=\left\{ \ba{rl} 1,&u>0
\\  -1,&u<0
\\ \ea \right\}
$, and ${\rm Tr}$ is the trace of an integro-differential operator
in the functional space: ${\rm Tr} \,\, U=\int d^3{\bf r}\,{\rm
tr}\langle{\bf r}|U|{\bf r}\rangle$. The thermal expectation value
of the observable is conventionally defined as (see, e.g.,
Ref.\cite{Das}) \be
O_\Upsilon(T)=\langle\hat{O}_\Upsilon\rangle_\beta\equiv \frac{{\rm
Sp}\,\hat{O}_\Upsilon \exp(-\beta\hat{O}_H)}{{\rm
Sp}\exp(-\beta\hat{O}_H)},\,\,\,\beta=(k_B T)^{-1}, \ee where
$T$ is the equilibrium temperature, $k_B$ is the Boltzmann constant,
and $\rm Sp$ is the trace or the sum over the expectation values in
the Fock state basis in the second-quantized theory. Evidently, the
zero-temperature limit of Eq.(18) coincides with Eq.(17): \be
O_\Upsilon(0)=\langle{\rm vac|\hat{O}_\Upsilon|{\rm vac}}\rangle.
\ee Thus, Eq.(18) can be presented in a way similar to that of
Eq.(17), i.e., through the functional trace of operators in the
first-quantized theory, see, e.g., Ref.\cite{Nie5}, \be
O_\Upsilon(T)=-\frac{1}{2}{\rm Tr}\Upsilon\tanh(\frac12\beta H). \ee
The self-adjointness of the Dirac hamiltonian ensures the
conservation of energy in the second-quantized theory, and the
corresponding operator is diagonal in creation and destruction
operators, \be \hat{O}_H=\sum\!\!\!\!\!\!\!\int
E_\lambda[a^+_\lambda a_\lambda-b^+_\lambda b_\lambda-\frac12{\rm
sgn}(E_\lambda)];\ee the operator of any other conserved observable
is diagonal as well.

If at least one of two observables is conserved, then their thermal
correlation, \be\Delta
(T;\,\,\hat{O}_{\Upsilon_1},\,\,\hat{O}_{\Upsilon_2})=\langle
\hat{O}_{\Upsilon_1}\,\,\hat{O}_{\Upsilon_2}\rangle_\beta-
\langle\hat{O}_{\Upsilon_1}\rangle_\beta\langle\hat{O}_{\Upsilon_2}\rangle_\beta,
\ee takes the form \be
\Delta(T;\,\,\hat{O}_{\Upsilon_1},\,\,\hat{O}_{\Upsilon_2})=\frac14{\rm
Tr}\Upsilon_1\Upsilon_2\,{\rm sech}^2(\frac12\beta H). \ee In
particular, the thermal quadratic fluctuation of a conserved
observable takes form \be
\Delta(T;\,\,\hat{O}_{\Upsilon},\,\,\hat{O}_{\Upsilon})=\frac14{\rm
Tr}\Upsilon^2\,{\rm sech}^2(\frac12\beta H). \ee

It is instructive to present Eqs.(20) and (23) in terms of contour
integrals involving the resolvent of the Dirac hamiltonian: \be
O_\Upsilon(T)=-\frac12\int\limits_{C}\frac{d\omega}{2\pi
i}\tanh(\frac12\beta\omega)\,{\rm Tr}\Upsilon(H-\omega)^{-1} \ee and
\be \Delta(T;\,\,\hat{O}_{\Upsilon_1},\,\,\hat{O}_{\Upsilon_2})=
\frac14\int\limits_{C}\frac{d\omega}{2\pi i}{\rm
sech}^2(\frac12\beta \omega)\,{\rm Tr}\Upsilon_1\Upsilon_2
(H-\omega)^{-1}, \ee where $C$ is the contour consisting of two
collinear straight lines, $(-\infty+i0,\,\,+\infty+i0)$ and
$(+\infty-i0,\,\,-\infty-i0)$, in the complex $\omega$-plane. Note
that only the even part of ${\rm Tr}\Upsilon(H-\omega)^{-1}$
contributes to thermal average $O_\Upsilon(T)$, and only the odd
part of ${\rm Tr}\Upsilon_1\Upsilon_2(H-\omega)^{-1}$ contributes to
thermal correlation
$\Delta(T;\,\hat{O}_{\Upsilon_1},\,\hat{O}_{\Upsilon_2})$. By
deforming contour $C$ to encircle poles of the
$\tanh(\frac12\beta\omega)$ and ${\rm sech}^2(\frac12\beta\omega)$
functions, which occur along the imaginary axis, one gets \be
O_\Upsilon(T)=-\frac{1}{\beta}\sum\limits_{n\in\mathbb{Z}}{\rm
Tr}\Upsilon(H-i\omega_n)^{-1} \ee and \be
\Delta(T;\,\,\hat{O}_{\Upsilon_1},\,\,\hat{O}_{\Upsilon_2})=
-\frac{1}{\beta^2}\sum\limits_{n\in\mathbb{Z}}{\rm
Tr}\Upsilon_1\Upsilon_2(H-i\omega_n)^{-2}, \ee where
$\omega_n=(2n+1)\pi/\beta$. Alternatively, by deforming contour $C$
around poles and cuts of the spectrum of $H$, which lie on the real
axis, one gets \be
O_\Upsilon(T)=-\frac{1}{2}\int\limits_{-\infty}^{\infty}d
E\,\tau_\Upsilon (E) \tanh(\frac12\beta E) \ee and \be
\Delta(T;\,\,\hat{O}_{\Upsilon_1},\,\,\hat{O}_{\Upsilon_2})=
\frac{1}{4}\int\limits_{-\infty}^{\infty}d
E\,\tau_{\Upsilon_1\Upsilon_2}(E) \,{\rm sech}^2(\frac12\beta E),
\ee where \be \tau_\Upsilon(E)=\pm\frac1\pi I m{\rm
Tr}\Upsilon(H-E\mp i0)^{-1} \ee and \be
\tau_{\Upsilon_1\Upsilon_2}(E)=\pm \frac1\pi I m{\rm
Tr}\Upsilon_1\Upsilon_2(H-E\mp i0)^{-1} \ee are the appropriate
spectral densities. Expression (29) and (30) can be regarded as the
Sommerfeld--Watson transforms of the infinite sum expressions,
Eqs.(27) and (28). Note that only the odd part of $\tau_\Upsilon(E)$
contributes to $O_\Upsilon(T)$ and only the even part of
$\tau_{\Upsilon_1 \Upsilon_2}(E)$ contributes to
$\Delta(T;\,\hat{O}_{\Upsilon_1},\,\hat{O}_{\Upsilon_2})$.

The Dirac hamiltonian in the background of a static magnetic
monopole takes form \be H=-\balpha\cdot(\i\bpartial+e{\bf A})+\gamma^0 M,
\ee where $\balpha=\gamma^0\bgamma$, and $\gamma^0$, $\bgamma$ are
the Dirac matrices, $M$ is the electron mass, and ${\bf A}$ is given
by Eq.(2). The magnetic monopole background is rotationally
invariant and three generators of rotations are identified with the
components of vector ${\bf J}$ -- the operator of total angular
momentum in the first-quantized theory, \be {\bf J}=\bLambda +
\bSigma,\ee where \be \bLambda=-{\bf r}\times(i\bpartial+e{\bf
A})-eg\frac{\bf r}{r} \ee is its orbital part, and \be \bSigma
=\frac{1}{4i}\balpha\times \balpha\ee is its spin part; note that
the last term in Eq.(35) is necessary in order to ensure the correct
commutation relations:
$$
[J^j,\,J^k]_-=i\varepsilon^{jkl}J^l.
$$

The nonvanishing of any
component of the vector vacuum expectation value in the
second-quantized theory,
$$
O_{\bf J}(0)=-\frac12{\rm Tr}{\bf
J}\,{\rm sgn}(H),
$$
would point at spontaneous breaking of the rotational invariance
(nonuniqueness of the ground state). Even if $O_{\bf J}(0)$ is
vanishing, it may happen that quantity \be O_{{\bf
J}^2}(0)=-\frac12{\rm Tr}{\bf J}^2 \,{\rm sgn}(H) \ee is
nonvanishing, which is compatible with the uniqueness of the ground
state preserving the rotational invariance in the second-quantized
theory. Note also that in the first-quantized theory operators $H$,
${\bf J}^2$ and any component of vector ${\bf J}$ are commuting,
therefore the corresponding operators in the second-quantized theory
can be diagonalized. On the contrary, operators $\hat{O}_{{\bf
\scriptstyle\Lambda}^2}$ and $\hat{O}_{{\bf \scriptstyle\Sigma}^2}$
are not diagonalizable, and, consequently, quantities \be O_{{\bf
\scriptstyle\Lambda}^2}(0)=-\frac12{\rm Tr}\bLambda^2{\rm sgn}(H)
\ee and \be O_{{\bf \scriptstyle\Sigma}^2}(0)=-\frac12{\rm
Tr}\bSigma^2{\rm sgn}(H) \ee have to be regarded as vacuum averages
rather than sharp quantum observables. As to quantity $O_{{\bf
J}^2}(0)$ (37), one might anticipate that it is a sharp observable,
which is substantiated by the fact that its thermal quadratic
fluctuation, \be \Delta(T;\,\hat{O}_{{\bf J}^2},\,\hat{O}_{{\bf
J}^2})=\frac14{\rm Tr}{\bf J}^4\,{\rm sech}(\frac14\beta H), \ee
tends to zero in the limit $T\rightarrow0$
($\beta\rightarrow\infty$). However, we shall find special
circumstances when the fluctuation is nonzero at zero temperature
and squared total angular momentum is not a sharp observable even at
zero temperature.

\section{Solutions to the Dirac equation in the monopole background}

The usual spherical harmonics $Y_{lm}(\vartheta,\phi)$ are
replaced by the (two-valued with two different values in $R_a$
and $R_b$ --- see Section I) monopole harmonics
$Y_{q,l,m}(\vartheta,\phi)$ \cite{WY2}, since orbital angular momentum,
see Eq.(35), differs from the usual one:
\be
\label{SPH-G}
\begin{array}{c}
Y_{q,l,m}(\vartheta,\phi) = M_{qlm} (1-\cos\vartheta)^{\frac\alpha2}
(1+\cos\vartheta)^{\frac\beta2}
P^{\alpha,\beta}_{l+m}(\cos\vartheta)\, \e^{\i(m\pm q)\phi},\\ q=eg,\quad
\alpha=-q-m,\quad \beta=q-m, \\
M_{qlm}=2^m \sqrt{\frac {(2l+1)(l-m)!(l+m)!}
{4\pi(l-q)!(l+q)!}},\quad l=|q|,|q|+1,\ldots,\quad m=-l,\,-l+1,\,\ldots,\, l,
\end{array}
\ee
where
$$
P^{\alpha,\beta}_n(u)=\frac{(-1)^n}{2^n n!}
(1-u)^{-\alpha}(1+u)^{-\beta}
\frac{d^n}{du^n}[(1-u)^{\alpha+n}(1+u)^{\beta+n}]
$$
are the Jacobi polynomials, see, e.g., Ref.\cite{Abra}. The plus
sign is chosen for $R_a$ and the minus sign is chosen for $R_b$. The
nontrivial nature of wave functions is completely embedded in the
monopole harmonics.

The eigensections of ${\bf J}^2$ and $J_z$ with eigenvalues equal to
$j(j+1)$ and $m$ correspondingly are \cite{Kaz}
\be \label{phi}
\varphi^{(1)}_{jm}(\vartheta,\phi)= \left(
\begin{array}{c}\sqrt{\frac{j+m}{2j}}
Y_{q,j-\frac12,\,m-\frac12}(\vartheta,\phi)\\ \sqrt{\frac{j-m}{2j}}
Y_{q,j-\frac12,\,m+\frac12}(\vartheta,\phi)
\end{array}
\right),\quad \varphi^{(2)}_{jm}(\vartheta,\phi)=
\left(\begin{array}{c}-\sqrt{\frac{j-m+1}{2j+2}}
Y_{q,j+\frac12,\,m-\frac12}(\vartheta,\phi)\\ \sqrt{\frac{j+m+1}{2j+2}}
Y_{q,j+\frac12,\,m+\frac12}(\vartheta,\phi)
\end{array}\right),
\ee
where $j\geq|q|+\frac12$ for $\varphi_{jm}^{(1)}$, and $j\geq |q|-\frac12$
for $\varphi_{jm}^{(2)}$. One chooses the
following linear combinations in the case of $j\geq|q|+\frac12$:
$$
\xi^{(1)}_{jm}(\vartheta,\phi) = c_j\varphi^{(1)}_{jm}(\vartheta,\phi)
- s_j\varphi^{(2)}_{jm}(\vartheta,\phi),\quad
\xi^{(2)}_{jm}(\vartheta,\phi) = s_j\varphi^{(1)}_{jm}(\vartheta,\phi)
+ c_j\varphi^{(2)}_{jm}(\vartheta,\phi),
$$
\be
c_j=\frac{\sgn(q)(\sqrt{2j+1+2q}+\sqrt{2j+1-2q})}
{2\sqrt{2j+1}},
\quad
 s_j=\frac{\sgn(q)(\sqrt{2j+1+2q}-\sqrt{2j+1-2q})}
{2\sqrt{2j+1}},
\ee
which satisfy the system of equations
\be
\ba{c}
-\bsigma(\i\bpartial+e{\bf A})h(r)\xi^{(1)}_{jm}=
\i (\partial_r + r^{-1}- \mu r^{-1})h(r)\xi^{(2)}_{jm}\\
-\bsigma(\i\bpartial+e{\bf A})h(r)\xi^{(2)}_{jm}= \i
(\partial_r+ r^{-1}+ \mu r^{-1})h(r)\xi^{(1)}_{jm}
\ea
\ee
for an arbitrary $h(r)$; here and in the following $\sigma^j$ are
the Pauli matrices and
\be
\mu=\sqrt{\left(j+\frac12\right)^2-q^2} \,.
\ee
In the case of $j=|q|-\frac12$ one defines
\be
\label{eta}
\eta_m(\vartheta,\phi)=\varphi^{(2)}_{jm}(\vartheta,\phi),
\ee
which satisfies
\be \bsigma(\i\bpartial+e{\bf A})h(r)\eta_m=
\i\sgn(q) (\partial_r+r^{-1})h(r)\eta_m.
\ee

In the standard representation for the Dirac matrices,
$$
\gamma^0=\left(
\ba{cc} 1 & 0\\
0 & -1\ea\right),\; \balpha=\left(
\ba{cc}
0 & \bsigma \\
\bsigma & 0
\ea
\right),
$$
the spin part of angular momentum (36) is of the block-diagonal form,
\be
{\bf \Sigma}=\frac12\left(
\ba{cc} \bsigma & 0\\
0 & \bsigma \ea\right),
\ee
and the solutions to the Dirac equation (\ref{Dirac_equation}) in the
monopole background are constructed as follows,\\
type 1 $(j\geq|q|+\frac12)$:
\be
\la{\bf r}|E,j,m\ra^{(1)} = \left(
\begin{array}{c}
\sqrt{1+\frac{M}{E}}\, \sqrt{\frac{k}{2r}}\,
J_{\mu-\frac12}(kr)\, \xi^{(1)}_{jm}(\vartheta,\phi)
\\
-\i \sgn (E)\, \sqrt{1-\frac{M}{E}}\, \sqrt{\frac{k}{2r}}\,
J_{\mu+\frac12}(kr)\, \xi^{(2)}_{jm}(\vartheta,\phi)
\end{array}
\right)\,,
\ee
type 2 $(j\geq|q|+\frac12)$:
\be
\la{\bf r}|E,j,m\ra^{(2)} = \left(
\begin{array}{c}
\sqrt{1+\frac{M}{E}}\,\sqrt{\frac{k}{2r}}\,
J_{\mu+\frac12}(kr)\, \xi^{(2)}_{jm}(\vartheta,\phi)
\\
\i \sgn (E)\, \sqrt{1-\frac{M}{E}}\, \sqrt{\frac{k}{2r}}\,
J_{\mu-\frac12}(kr)\, \xi^{(1)}_{jm}(\vartheta,\phi)
\end{array}
\right)\,,
\ee
type 3 $(j=|q|-\frac12)$:
\be
\label{type_3}
\la{\bf r}|E,m\ra_\Theta= \left(
\begin{array}{c}
f(r)\eta_m(\vartheta,\phi)
\\
g(r)\eta_m(\vartheta,\phi)
\end{array}
\right)\,, \ee where $k=\sqrt{E^2-m^2}$, $J_\rho(u)$ is the Bessel
function of order $\rho$, and radial functions $f(r)$ and $g(r)$ are
divergent, although square integrable, at the origin. That is why
the type 3 solution is called irregular, in contrast to the types 1
and 2 solutions which are regular at the origin. The procedure of
the self-adjoint extension is implemented for the partial
hamiltonian with $j=|q|-\frac12$, yielding the boundary condition
for the corresponding partial mode \cite{Gold}: \be \ba{c}
\cos\left(\frac\Theta 2+\frac\pi 4\right)\lim\limits_{r\rightarrow
0} r f(r)=\i\sgn (q) \sin\left(\frac\Theta 2+\frac\pi
4\right)\lim\limits_{r\rightarrow 0} r g(r), \ea \ee where $\Theta$
is the self-adjoint extension parameter. This gives the explicit
form for the radial functions in Eq.(51) \cite{Yam} \be \ba{c} f(r)=
\frac{\i\,\sgn(q)}{r\sqrt{\pi E(E-M\sin\Theta)}} \left[ (E+M)
\sin{kr}\, \cos\left(\frac\Theta2+\frac\pi4\right)
+k\cos kr\,\sin\left(\frac\Theta2+\frac\pi4\right)\right], \\
g(r)=\frac{1}{r\sqrt{\pi E(E-M\sin\Theta)}} \left[k\cos kr\,
\cos\left(\frac\Theta2+\frac\pi4\right) -(E-M)\sin kr\,
\sin\left(\frac\Theta2+\frac\pi4\right)\right].
\ea
\ee
If $\cos\Theta<0$, then there exists in addition a bound state with energy $E_{BS} =
M\sin\Theta$:
\be
\la{\bf r}|E_{BS},m\ra_\Theta = \frac1r \left(
\begin{array}{c}
\i\sgn(q)\sin(\frac{\Theta}{2}+\frac{\pi}{4})\eta_m(\vartheta,\phi)\\
\cos(\frac{\Theta}2+\frac{\pi}4)\,\eta_m(\vartheta,\phi)
\end{array}
\right) \sqrt{-2M\cos\Theta}\, \e^{Mr\cos\Theta}. \ee These
solutions form a complete orthonormalized set: \be
\begin{array}{r}
{}^{(i)}\langle E,j,m|E',j',m'\rangle^{(i')} = \delta_{ii'}\,
\delta_{jj'}\, \delta_{mm'}\,
\delta_{\sgn(E),\sgn(E')}\, \delta(k-k')\,,\quad i,\,i'=1,2,
\\
{}_\Theta\langle E,m|E',m'\rangle_\Theta = \delta_{mm'}\,
\delta_{\sgn(E),\sgn(E')}\, \delta(k-k')\,,
\\
{}_\Theta\langle E_{BS},m|E_{BS},m'\rangle_\Theta = \delta_{mm'} \,.
\end{array}
\ee

\section{Induced quantum numbers at zero temperature}

The vacuum expectation values are calculated using formula (17) and the
classical solutions from the previous section. Let us consider an observable
which corresponds to an operator in the first-quantized theory of
the block-diagonal form,
\be \label{block-diagonal} \Upsilon=\left(
\begin{array}{cc}
\Omega & 0\\
0 & \Omega
\end{array}
\right) ,\ee and containing no derivatives in $r$: \be \label{block}
\Omega\,
h(r)\varphi(\vartheta,\phi)=h(r)\Omega\,\varphi(\vartheta,\phi). \ee
For the contribution of the type 1 solutions, one gets \be
{}^{(1)}\la E,j,m|{\bf r}\ra \Upsilon \la{\bf r}|E,j,m\ra^{(1)} =
\frac{k}{2r} \left[ (1+\frac{M}{E}) J^2_{\mu-\frac12}(kr)
\xi^{(1)\dagger}_{jm}\Omega\xi^{(1)}_{jm} + (1-\frac{M}{E})
J^2_{\mu+\frac12}(kr) \xi^{(2)\dagger}_{jm}\Omega\xi^{(2)}_{jm}
\right]; \ee for that of the type 2 solutions, one gets  \be
{}^{(2)}\la E,j,m|{\bf r}\ra \Upsilon \la{\bf r}|E,j,m\ra^{(2)} =
\frac{k}{2r} \left[ (1+\frac{M}{E}) J^2_{\mu+\frac12}(kr)
\xi^{(2)\dagger}_{jm}\Omega\xi^{(2)}_{jm} + (1-\frac{M}{E})
J^2_{\mu-\frac12}(kr) \xi^{(1)\dagger}_{jm}\Omega\xi^{(1)}_{jm}
\right]. \ee Summing Eqs.(58) and (59), one gets \be
\label{cancellation} \sum\limits_{i=1}^2{}^{(i)}\la E,j,m|{\bf r}\ra
\Upsilon \la{\bf
r}|E,j,m\ra^{(i)}=\frac{k}{r}\left[J^2_{\mu-\frac12}(kr)
\xi^{(1)\dagger}_{jm}\Omega\xi^{(1)}_{jm} +J^2_{\mu+\frac12}(kr)
\xi^{(2)\dagger}_{jm}\Omega\xi^{(2)}_{jm}\right], \ee which is
independent of the sign of $E$; thus, the overall contribution of
the types 1 and 2 solutions to Eq.(17) is zero. For the contribution
of the type 3 solution (continuous spectrum) to Eq.(17), one gets a
nonzero result: \be \label{type3_cont_spec}
-\frac12\sum\limits_{\sgn(E)} {}_{\Theta}\la E,m|{\bf r}\ra \Upsilon
\la{\bf r}|E,m\ra_{\Theta}\sgn (E) =
\frac{2\eta^\dagger_m\Omega\eta_m kM\sin\Theta} {\pi r^2|E|
(k^2+M^2\cos^2\Theta)} (k\cos2kr+M\sin2kr\cos\Theta). \ee In order
to perform integration over $k$ one can take parity into account and
extend integration from $(0,\infty)$ to $(-\infty,\infty)$,
replacing $k\sin kr\to -\i k\e^{\i kr},$ $\cos kr\to \e^{\i kr}$.
Adding the contribution of the bound state and summing over $m$, one
gets
\be \ba{lcr}-\frac12\tr\la{\bf r}|\Upsilon\sgn (H)|{\bf r}\ra
& = & \displaystyle - \frac{M}{2r^2} \sum\limits_{m=-|q|}^{|q|}
\eta^\dagger_m\Omega\eta_m\, \Bigl\{ \cos\Theta \bigl[ \sgn(\sin 2\Theta) -
\sgn(\sin\Theta) \bigr] \e^{2Mr\cos\Theta}
\\
& & \displaystyle +\frac{\sin\Theta}{\pi}
\int\limits_{-\infty}^\infty \! dk\, \frac{k \e^{2\i
kr}}{\sqrt{k^2+M^2}(k+\i M\cos\Theta)} \Bigr\}.
\ea
\ee
The contour of integration can be deformed to the upper half-plane of complex $k$ to
enclose a cut along the imaginary axis at $Im\,k>M$ and encircle a pole occuring
in the case of $\cos\Theta<0$ at $Im\,k=-M\cos\Theta$. The contribution of the
pole cancels that of the bound state, and only the contribution from
the cut survives. Averaging over the angular variables yields
\be
\begin{array}{r}
\displaystyle \rho_{\Upsilon}(r)= \frac{1}{4\pi}\int\limits_{0}^{2\pi}
d\phi\int\limits_{0}^{\pi}d\vartheta\sin\vartheta\,(-\frac12){\rm tr}\la{\bf r}
|\Upsilon{\rm sgn}(H)|{\bf r}\ra=
\\
\displaystyle =- \sum\limits_{m=-|q|}^{|q|} \int\limits_{0}^{2\pi}d
\phi\int\limits_{0}^{\pi}d\vartheta\sin\vartheta\,\,
\eta^\dagger_m\Omega\eta_m \frac{M\sin\Theta}{(2\pi)^2 r^2}
\int\limits_M^\infty{\!d\kappa\, \frac{\kappa\e^{-2\kappa r}}
{\sqrt{\kappa^2-M^2}\,(\kappa+M\cos\Theta)}},
\end{array}
\ee and the vacuum expectation value takes form \be
\label{total_average} O_\Upsilon(0) = 4\pi\int\limits_0^\infty\! dr\, r^2
\rho_{\Upsilon}(r) = -\sum_{m=-|q|}^{|q|}\int\limits_{0}^{2\pi}\!d\phi\int\limits_0^\pi
d\vartheta\sin\vartheta\,\eta^\dagger_m\Omega\eta_m \,\,
\frac1\pi\arctan\left(\tan\frac\Theta2\right).
\ee

In the case of $\Omega=I$, where $I$ is the $2\times2$ unity matrix,
using the orthonormality of ${\eta_m}^,s$, one gets \be
\sum\limits_{m=-|q|}^{|q|}\int\limits_{0}^{2\pi}d\phi\int\limits_{0}^{\pi}
d\vartheta \sin\vartheta\,\eta^\dagger_m\eta_m =2|q|, \ee and the
vacuum expectation value of fermion number takes form \cite{Gro,Yam}
\be O_I(0)=-2|eg|\,\frac1\pi\arctan\left(\tan\frac\Theta2\right),
\ee where we have recalled relation $q=eg$. Multiplying Eq.(66) by
$e$, one gets the induced vacuum charge (11). Note that Eq.(66) at
$|eg|=\frac12$ coincides with the expression for the fermion number
which is induced in $2+1$-dimensional space-time in the vacuum by a
pointlike magnetic vortex with flux $\pi\mod2\pi$ \cite{Sit1}.

It is straightforward to prove that angular momentum, as well as its spin and
orbital parts separately, is not induced in the vacuum, and, consequently,
rotational invariance is not spontaneously broken. Indeed, since $J_z\eta_m=m\eta_m$,
one gets
\be
O_{J_z}(0)=0
\ee
due to summation over $m$. As to the $z$-component of orbital angular momentum,
this issue is more intricate. Using Eqs.(42) and (46) and relation
$\Lambda_zY_{q,l,m}=mY_{q,l,m}$, one gets
$$
\eta^\dagger_m\Lambda_z\eta_m=\left(2|q|+1\right)^{-1}\left[\left(|q|+\frac12-m\right)
\left(m-\frac12\right)\left|Y_{q,|q|,m-\frac12}\right|^2+\right.
$$
$$
\left.+\left(|q|+\frac12+m\right)
(m+\frac12)\left|Y_{q,|q|,m+\frac12}\right|^2\right].
$$
Intergating over angular variables yields:
$$
\int\limits_{0}^{2\pi}d\phi\int\limits_{0}^{\pi}d\vartheta\,
\sin\vartheta\,\eta^\dagger_m\Lambda_z\eta_m=
\frac{2(|q|+1)}{2|q|+1}m,
$$
where the orthonormality of the monopole harmonics has been used. Summation over $m$
results in zero, and the same considerations apply to the $z$-component of spin:
\be
O_{\Lambda_z}(0)=O_{\Sigma_z}(0)=0.
\ee
Similarly, one can show that components $O_{J_x\pm iJ_y}(0)$, $O_{\Lambda_x\pm i\Lambda_y}(0)$,
$O_{\Sigma_x\pm i\Sigma_y}(0)$ vanish. Here again the orthonormality is crucial:
roughly speaking, the diagonal matrix elements of raising and lowering operators are
equal to zero.

Let us turn now to the vacuum expectation values of the squares of
orbital angular momentum, spin, and total angular momentum. Using
relation ${\bf \Lambda}^2Y_{q,l,m} =l(l+1)Y_{q,l,m}$, where $l=|q|$,
one gets \be
\sum\limits_{m=-|q|}^{|q|}\int\limits_{0}^{2\pi}d\phi\int\limits_{0}^{\pi}
d\vartheta\,\sin\vartheta\,\eta^\dagger_m{\bf
\Lambda}^2\eta_m=2q^2(|q|+1). \ee Taking account for relation
$\frac14{\bsigma}^2=\frac34I$, one gets immediately \be
\sum\limits_{m=-|q|}^{|q|}\int\limits_{0}^{2\pi}d\phi\int
\limits_{0}^{\pi}d\vartheta\,\sin\vartheta\,
\eta^\dagger_m\frac14\bsigma^2\eta_m=\frac32|q|. \ee Using relation
${\bf J}^2\eta_m=j(j+1)\eta_m$, where $j=|q|-\frac12$, one gets \be
\sum\limits_{m=-|q|}^{|q|}\int\limits_{0}^{2\pi}d\phi
\int\limits_{0}^{\pi}d\vartheta\,\sin\vartheta\, \eta^\dagger_m{\bf
J}^2\eta_m=2|q|\left(q^2-\frac14\right). \ee Thus, we get
following expressions for the vacuum expectation values:
\be
O_{{\bf \Lambda}^2}(0)=|eg|(|eg|+1)O_I(0),
\ee
\be O_{{\bf \Sigma}^2}(0)=\frac34
O_I(0),
\ee
\be O_{{\bf J}^2}(0)=\left[(eg)^2-\frac14\right]O_I(0), \ee
where induced vacuum fermion number $O_I(0)$ is given by Eq.(66).

\section{Spectral densities}

An alternative and more refined way of treating the induced quantum
numbers, which is especially adapted to the case of nonzero temperature,
involves the use of spectral densities, see Eqs.(31) and (32). In general,
the spectral density is decomposed as
\be
\tau_\Upsilon(E)=\tau_\Upsilon^{(0)}(E)+\tau_\Upsilon^{\rm ren}(E),
\ee
where
\be
\tau_\Upsilon^{(0)}(E)=\pm \frac1\pi Im {\rm Tr}\,\Upsilon^{(0)}(H^{(0)}-E\mp i0)^{-1}
\ee
is the spectral density in the absence of interaction (the operators in
the free first-quantized theory are denoted by $H^{(0)}$ and $\Upsilon^{(0)}$), and
\be
\tau_\Upsilon^{\rm ren}(E)=\pm \frac1\pi Im\left[{\rm Tr}\,\Upsilon(H-E\mp i0)^{-1}\right]_{\rm ren}
\ee
is the addition which is due to interaction with the background field;
the subscript $^{\rm ren}$ in the right hand side of Eq.(77) denotes the
renormalization of the functional trace by subtraction:
\be
\left[{\rm Tr}\Upsilon(H-\omega)^{-1}\right]_{\rm ren}=
{\rm Tr}\Upsilon(H-\omega)^{-1}-{\rm Tr}\Upsilon^{(0)}\left(H^{(0)}-\omega\right)^{-1}.
\ee

To compute $\tau_\Upsilon^{(0)}(E)$, let us consider matrix element
$$
\left\langle{\bf r}|\Upsilon^{(0)}(H^{(0)}\!-\!\omega)^{-1}|{\bf r}'\right\rangle =\Upsilon^{(0)}
\!\int\!\frac{d^3p}{(2\pi)^3}\,e^{i{\bf p}({\bf r}-{\bf r}')}\,
\frac{\balpha\cdot {\bf p}+\gamma^0M+\omega}{{\bf p}^2-\omega^2+M^2}.
$$
Although the integral in the right hand side of the last equation is
divergent, its imaginary part at $\omega=E\pm i0$ in the case of
${\bf r}'={\bf r}$ is finite: \be \pm Im\left\langle{\bf
r}|\Upsilon^{(0)}(H^{(0)}-E\mp i0)^{-1}|{\bf r}\right\rangle = \pi
\sgn(E)\int\frac{d^3p}{(2\pi)^3}\delta({\bf
p}^2-E^2+M^2)\Upsilon^{(0)} \left(\balpha\cdot{\bf
p}+\gamma^0M+E\right). \ee Taking $\bf \Sigma$ (36) and ${\bf
\Lambda}^{(0)}=-i{\bf r}\times \bpartial$ in the capacity of
$\Upsilon^{(0)}$, one gets immediately
$$
\pm Im {\rm tr}\left\langle{\bf r}|{\bf \Sigma}(H^{(0)}-E\mp i0)^{-1}|{\bf r}\right\rangle =
\pm Im{\rm tr}\left\langle{\bf r}|{\bf \Lambda}^{(0)}(H^{(0)}-E\mp i0)^{-1}|{\bf r}\right\rangle=0,
$$
and, consequently,
\be
\tau_{\bf \Sigma}^{(0)}(E)=\tau_{\bf \Lambda}^{(0)}(E)=\tau_{\bf J}^{(0)}(E)=0.
\ee
Taking ${\bf \Sigma}^2$ and $({\bf \Lambda}^{(0)})^2$, one gets nonzero results
$$
\pm Im {\rm tr}\left\langle{\bf r}|{\bf \Sigma}^2(H^{(0)}-E\mp i0)^{-1}|{\bf r}\right\rangle =
4\pi\sgn(E)\int\frac{d^3p}{(2\pi)^3}\delta(p^2-E^2+M^2)\frac34E=
$$
$$
=\frac{3}{4\pi}|E|(E^2-M^2)^{\frac12}\theta(E^2-M^2),
$$
$$
\pm Im {\rm tr}\left\langle{\bf r}|({\bf \Lambda}^{(0)})^2(H^{(0)}-E\mp i0)^{-1}|{\bf r}\right\rangle =
4\pi\sgn(E)\int\frac{d^3p}{(2\pi)^3}\delta(p^2-E^2+M^2)({\bf r}\times{\bf p})^2E=
$$
$$
=\frac{2}{3\pi}|E|r^2(E^2-M^2)^{\frac32}\theta(E^2-M^2),
$$
where $\theta(u)=\frac12[1+\sgn(u)]$. Consequently, we get \be
\tau_{{\bf
\Sigma}^2}^{(0)}(E)=\frac{3}{4\pi^2}V|E|(E^2-M^2)^{\frac12}\theta(E^2-M^2),
\ee \be \tau_{{\bf
\Lambda}^2}^{(0)}(E)=\frac{2}{5\pi^2}\left(\frac{3}{4\pi}\right)^{\frac23}
V^{\frac53}|E|(E^2-M^2)^{\frac32}\theta(E^2-M^2), \ee \be \tau_{{\bf
J}^2}^{(0)}(E)=\frac{V}{\pi^2}\left[\frac25\left(\frac{3V}{4\pi}\right)^{\frac23}
(E^2-M^2)+\frac34\right]|E|(E^2-M^2)^{\frac12}\theta(E^2-M^2), \ee
where $V=\int\limits_{0}^{2\pi}d\phi\int\limits_0^\pi d\vartheta\sin
\vartheta \int\limits_0^Rdrr^2$ is the volume of the spherical box
of radius $R$. In a similar way, one can get \be
\tau_{I}^{(0)}(E)=\frac{V}{\pi^2}|E|(E^2-M^2)^{\frac12}\theta(E^2-M^2),
\ee \be \tau_{{\bf \Sigma}^2{\bf
J}^2}^{(0)}(E)=\frac{3V}{4\pi^2}\left[\frac25\left(\frac{3V}{4\pi}\right)^{\frac23}
(E^2-M^2)+\frac34\right]|E|(E^2-M^2)^{\frac12}\theta(E^2-M^2), \ee
\be \tau_{{\bf \Lambda}^2{\bf
J}^2}^{(0)}(E)=\frac{V}{5\pi^2}\left[\frac87\left(\frac{3V}{4\pi}\right)^{\frac43}
(E^2-M^2)+\frac32\left(\frac{3V}{4\pi}\right)^{\frac23}\right]|E|(E^2-M^2)^{\frac32}\theta(E^2-M^2),
\ee \be \tau_{{\bf
J}^4}^{(0)}(E)=\frac{V}{\pi^2}\left[\frac{8}{35}\left(\frac{3V}{4\pi}\right)^{\frac43}
(E^2-M^2)^2+\left(\frac{3V}{4\pi}\right)^{\frac23}(E^2-M^2)+\frac{9}{16}\right]|E|(E^2-M^2)^{\frac12}\theta(E^2-M^2).
\ee It should be noted that Eqs.(81)-(87) are even in $E$, and,
thus, they do not contribute to the expectation values, while
contributing to the appropriate correlations and quadratic
fluctuations.

Let us turn now to the part of the spectral density, Eq.(77), which
is due to interaction with the monopole background, and decompose it
in the following way: \be \tau_\Upsilon^{\rm
ren}(E)=\tau_\Upsilon^{{\rm ren}'}(E)+\tau_\Upsilon^{(3)}(E), \ee
where $\tau_\Upsilon^{(3)}(E)$ is taking account of only the
contribution of the type 3 solutions, Eqs.(51), (53) and (54), while
$\tau_\Upsilon^{{\rm ren}'}(E)$ is including the contribution of the
types 1 and 2 solutions and subtracted plane wave solutions. In the
case of $\Upsilon$ in the block-diagonal form (56) with no
derivatives in $r$ (57), the total contribution of the types 1 and 2
solutions is even in $E$, see Eq.(60), and, thus, one gets \be
\tau_\Upsilon^{{\rm ren}'}(E)=\tau_\Upsilon^{{\rm ren}'}(-E). \ee As
to the contribution of the type 3 solutions, one obtains, following
Ref.\cite{Dun}, their contribution to the trace of resolvent \be
\left[{\rm
Tr}\Upsilon(H-\omega)^{-1}\right]_{(3)}=-\frac12\sum\limits_{m=-|q|}^{|q|}
\int\limits_{0}^{2\pi}d\phi\int\limits_{0}^{\pi}d\vartheta\sin\vartheta\,\eta_m^\dag
\Omega\eta_mM\frac{\omega\sin\Theta-M-i\sqrt{\omega^2-M^2}\cos\Theta}{(\omega^2-M^2)(\omega-M\sin\Theta)},
\ee where a physical sheet for square root is chosen as $0<{\rm
Arg}\sqrt{\omega^2-m^2}<\pi$ $(Im\sqrt{\omega^2-m^2}>0)$.
Consequently, we get \be \ba{lcr}
\tau_\Upsilon^{(3)}(E)&=&\sum\limits_{m=-|q|}^{|q|}\int\limits_{0}^{2\pi}d\phi
\int\limits_{0}^{\pi}d\vartheta\sin\vartheta\,\eta_m^\dag\Omega\eta_m
\left[\theta(-\cos\Theta)\delta(E-M\sin\Theta)-\frac14\delta(E-M)-\right. \\
& & \displaystyle \left.-\frac14\delta(E+M)+\frac{\cos\Theta}{2\pi}\frac{\sgn(E)}{E-M\sin\Theta}
\frac{M}{\sqrt{E^2-M^2}}\theta(E^2-M^2)\right].
\ea
\ee
One can conclude that irregular modes contribute both to expectation values and
to correlations and fluctuations, and their contribution is finite in the infinite
volume limit. On the contrary, the ideal gas (i.e. plane waves) contribution to
correlations and fluctuations diverges by power law as $V\rightarrow\infty$,
see Eqs.(81)-(87).

Using Eq.(91), one gets the following expression for the vacuum expectation value,
\be \ba{lcr}
O_\Upsilon(0)&=&-\frac12\sum\limits_{m=-|q|}^{|q|}\int\limits_{0}^{2\pi}d\phi
\int\limits_{0}^{\pi}d\vartheta\sin\vartheta\,\eta_m^\dag\Omega\eta_m\left[
\theta(-\cos \Theta)\sgn(\sin\Theta)+\right. \\ & & \displaystyle \left.
+\frac{\sin 2\Theta}{2\pi}\int\limits_{1}^{\infty}
\frac{dw}{\sqrt{w^2-1}}\frac{1}{w^2-\sin^2\Theta}\right],
\ea
\ee
which, after performing integration, yields Eq.(64), as it should be expected.

\section{Induced quantum numbers at nonzero temperature, \\
thermal correlations and quadratic fluctuations}

Using the results of the preceding section, we get the following
expression for the thermal expectation value (29), compare with
Eq.(92): \be \ba{lcr}
O_\Upsilon(T)=&-&\frac12\sum\limits_{m=-|q|}^{|q|}\int\limits_{0}^{2\pi}d\phi
\int\limits_{0}^{\pi}d\vartheta\sin\vartheta\,\eta_m^\dag\Omega\eta_m\left[
\theta(-\cos \Theta)\tanh(\frac12\beta M\sin\Theta)+\right. \\ & &
\displaystyle \left.+\frac{\sin
2\Theta}{2\pi}\int\limits_{1}^{\infty}
\frac{dw}{\sqrt{w^2-1}}\frac{\tanh(\frac12\beta
Mw)}{w^2-\sin^2\Theta}\right]. \ea \ee Taking the inverse
Sommerfeld-Watson transformation, see Eq.(27), we get the infinite
sum representation of Eq.(93): \be \ba{lcl}
O_\Upsilon(T)&=&-\sum\limits_{m=-|q|}^{|q|}\int\limits_{0}^{2\pi}d\phi
\int\limits_{0}^{\pi}d\vartheta\sin\vartheta\,\eta_m^\dag\Omega\eta_m\beta M\sin\Theta\times \\
 \displaystyle
& &\times
\sum\limits_{n=0}^{\infty}\left[(2n+1)^2\pi^2+\beta^2M^2+\beta
M\cos\Theta \sqrt{(2n+1)^2\pi^2+\beta^2M^2}\right]^{-1}. \ea \ee In
the case of $\Upsilon=I$, one gets induced fermion number
\cite{Cor,Dun}
\begin{eqnarray}
  \!\!\!\!\!\!\!\!\!\!O_I(T)&\!\!=&\!\!-|e
g|\left[\theta(-\cos\Theta)\tanh(\frac12\beta M\sin\Theta)+
\frac{\sin 2\Theta}{2\pi}\int\limits_{1}^{\infty}\frac{d
w}{\sqrt{w^2-1}}
\frac{\tanh(\frac12\beta M w)}{w^2-\sin^2\Theta}\right]\! =  \nonumber \\
 \displaystyle
 & \!\!=&\!\!-2|eg|\beta M\sin\Theta \sum\limits_{n=0}^{\infty}\left[(2n+1)^2\pi^2+\beta^2M^2+\beta M\cos\Theta
\sqrt{(2n+1)^2\pi^2+\beta^2M^2}\right]^{-1};
\end{eqnarray}
note that Eq.(95) at $|eg|=\frac12$ coincides with the expression
for fermion number which is induced in $2+1$-dimensional space-time
at nonzero temperature by a pointlike magnetic vortex with flux
$\pi\mod 2\pi$ \cite{Sit2}.

All other quantum numbers are related to Eq.(95): squared orbital angular momentum,
\be
O_{{\bf \Lambda}^2}(T)=|eg|(|eg|+1)O_I(T),
\ee
squared spin,
\be
O_{{\bf \Sigma}^2}(T)=\frac34O_I(T),
\ee
and squared total angular momentum
\be
O_{{\bf J}^2}(T)=\left[(eg)^2-\frac14\right]O_I(T);
\ee
incidentally, one gets
\be
O_{\bf \Lambda}(T)=O_{\bf \Sigma}(T)=O_{\bf J}(T)=0.
\ee

Let us turn now to thermal correlations and quadratic fluctuations
of observables. As it was shown in the previous section, the ideal
gas contribution (denoted by superscript $^{(0)}$) is prevailing
over the contribution (denoted by superscript $^{\rm ren}$) which is
due to interaction with the monopole background, since the former is
increasing, while the latter is constant as the volume of the system
increases. Using Eqs.(81)-(87), one gets:
\newline quadratic fluctuation of fermion number
\be
\Delta(T;\,\hat{O}_I,\,\hat{O}_I)=\frac{1}{4\pi^2}\frac{V}{\beta^3}\int\limits_{\beta^2M^2}^{\infty}
d s\frac{(s-\beta^2M^2)^\frac12}{\cosh^2(\frac12\sqrt{s})},
\ee
\newline correlation of fermion number and squared spin
\be
\Delta(T;\,\hat{O}_{{\bf \Sigma}^2},\,\hat{O}_I)=\frac{3}{16\pi^2}\frac{V}{\beta^3}\int\limits_{\beta^2M^2}^{\infty}
d s\frac{(s-\beta^2M^2)^\frac12}{\cosh^2(\frac12\sqrt{s})},
\ee
correlation of fermion number and squared orbital angular momentum
\be
\Delta(T;\,\hat{O}_{{\bf \Lambda}^2},\,\hat{O}_I)=\frac{1}{10\pi^2}
\left(\frac{3}{4\pi}\right)^{\frac23}\frac{V^\frac53}{\beta^5}\int\limits_{\beta^2M^2}^{\infty}
d s\frac{(s-\beta^2M^2)^\frac32}{\cosh^2(\frac12\sqrt{s})},
\ee
correlation of fermion number and squared total angular momentum
\be
\Delta(T;\,\hat{O}_{{\bf J}^2},\,\hat{O}_I)=\frac{1}{10\pi^2}
\left(\frac{3}{4\pi}\right)^{\frac23}\frac{V^\frac53}{\beta^5}\int\limits_{\beta^2M^2}^{\infty}
d s\frac{(s-\beta^2M^2)^\frac32}{\cosh^2(\frac12\sqrt{s})},
\ee
correlation of squared total angular momentum and squared spin
\be
\Delta(T;\,\hat{O}_{{\bf \Sigma}^2},\,\hat{O}_{{\bf J}^2})=\frac{3}{40\pi^2}
\left(\frac{3}{4\pi}\right)^{\frac23}\frac{V^\frac53}{\beta^5}\int\limits_{\beta^2M^2}^{\infty}
d s\frac{(s-\beta^2M^2)^\frac32}{\cosh^2(\frac12\sqrt{s})},
\ee
correlation of squared total angular momentum and squared orbital angular momentum
\be
\Delta(T;\,\hat{O}_{{\bf \Lambda}^2},\,\hat{O}_{{\bf J}^2})=\frac{2}{35\pi^2}
\left(\frac{3}{4\pi}\right)^{\frac43}\frac{V^\frac73}{\beta^7}\int\limits_{\beta^2M^2}^{\infty}
d s\frac{(s-\beta^2M^2)^\frac52}{\cosh^2(\frac12\sqrt{s})},
\ee
quadratic fluctuation of squared total angular momentum
\be
\Delta(T;\,\hat{O}_{{\bf J}^2},\,\hat{O}_{{\bf J}^2})=\frac{2}{35\pi^2}
\left(\frac{3}{4\pi}\right)^{\frac43}\frac{V^\frac73}{\beta^7}\int\limits_{\beta^2M^2}^{\infty}
d s\frac{(s-\beta^2M^2)^\frac52}{\cosh^2(\frac12\sqrt{s})},
\ee
where only the leading powers of volume in the large volume limit are retained.

In the high-temperature limit induced quantum numbers tend to zero as inverse temperature:
\be
O_\Upsilon(T\rightarrow\infty)=-\frac18\beta M\sin\Theta\sum\limits_{m=-|q|}^{|q|}
\int\limits_{0}^{2\pi}d\phi\int\limits_{0}^{\pi}d\vartheta\sin\vartheta\,\eta_m^\dag\Omega\eta_m,
\ee
whereas fluctuations and correlations increase as powers of temperature:
\be
\Delta(T\rightarrow\infty;\,\hat{O}_I,\,\hat{O}_I)=\frac43\Delta(T\rightarrow\infty;\,
\hat{O}_{{\bf \Sigma}^2},\,\hat{O}_I)=\frac13\frac{V}{\beta^3},
\ee
\be
\Delta(T\rightarrow\infty;\,\hat{O}_{{\bf \Lambda}^2},\,\hat{O}_I)=
\Delta(T\rightarrow\infty;\,\hat{O}_{\bf J^2},\,\hat{O}_I)=\frac43\Delta(T\rightarrow\infty;\,
\hat{O}_{{\bf \Sigma}^2},\,\hat{O}_{{\bf J}^2})=\frac{7\pi}{5^2}\left(\frac\pi 6\right)^{\frac13}
\frac{V^{\frac53}}{\beta^5},
\ee
\be
\Delta(T\rightarrow\infty;\,\hat{O}_{{\bf \Lambda}^2},\,\hat{O}_{{\bf J}^2})=
\Delta(T\rightarrow\infty;\,\hat{O}_{\bf J^2},\,\hat{O}_{\bf J^2})=\frac{31\pi^3}{5\cdot7^2}
\left(\frac6\pi\right)^{\frac13}
\frac{V^{\frac73}}{\beta^7}.
\ee

Thermal expectation value (93) can be presented as
\be
O_\Upsilon(T)=O_\Upsilon(0)+O_\Upsilon^{(\Delta)}(T),
\ee
where $O_\Upsilon(0)$ is given by Eq.(64), and
\begin{eqnarray}
O_\Upsilon^{(\Delta)}(T)&=&\sum\limits_{m=-|q|}^{|q|}\int\limits_{0}^{2\pi}
d\phi\int\limits_{0}^{\pi}d\vartheta\sin\vartheta\,\eta_m\Omega\eta_m
\left\{\frac{\theta(-\cos\Theta)\sgn_0(\sin\Theta)}{\exp(\beta M|\sin\Theta|)+1}\right.+ \nonumber \\
    &&\left.+\frac{\beta M}{4\pi}\int\limits_{1}^{\infty}dw\frac{\arctan\left[(1-w^{-2})^{\frac12}
    \tan \Theta\right]}{\cosh^2(\frac12\beta Mw)}\right\},
\end{eqnarray}
where \begin{math}{\rm sgn}_0(u)=\left\{\begin{array}{cc}
                                                       \sgn(u), & u\neq 0 \\
                                                       0, & u=0
                                                     \end{array}\right\}\end{math}.
One can verify that relation $\left.O_\Upsilon^{(\Delta)}(T)\right|_{\Theta=\pi\mod 2\pi}=0$
holds, and, thus, Eq.(112) vanishes exponentially in the zero-temperature limit (as $e^{-\beta M}$
at $\beta\rightarrow\infty$) for all values of $\Theta$.

At a first glance, one may anticipate that also thermal correlations and quadratic fluctuations vanish
exponentially in this limit for all values of $\Theta$, since the prevailing ideal gas
contribution is $\Theta$-independent. However, the bound state with zero energy
($E_{BS}=0$, i.e., $\Theta=\pi\mod 2\pi$, see Eq.(54)) in the one-particle spectrum reveals itself
in a completely different manner, as compared to Eq.(112). In the zero-temperature limit,
both the ideal gas contribution and the renormalized contribution of the types 1 and 2
solutions to correlations and fluctuations vanish exponentially, whereas the contribution of
the type 3 solutions behaves otherwise: the bound state pole in spectral density (91) is not
exponentially damped in
this limit if the bound state energy is zero. In general, we get
\be
\Delta(T\rightarrow0;\,\hat{O}_{\Upsilon_1},\,\hat{O}_{\Upsilon_2})=
\sum\limits_{m=-|q|}^{|q|}\int_{0}^{2\pi}d\phi\int\limits_{0}^{\pi}d\vartheta
\sin\vartheta\,\eta_m^\dag\Omega_1\Omega_2\eta_m\left\{\begin{array}{cc}
                                                       0, & \Theta\neq\pi\mod 2\pi \\
                                                       \frac14, & \Theta=\pi\mod 2\pi
                                                     \end{array}.\right.
\ee
In particular, the zero-temperature limits of the quadratic fluctuations of fermion number and
squared total angular momentum are
\be
\Delta(T\rightarrow0;\,\hat{O}_I,\,\hat{O}_I)=\left\{\begin{array}{cc}
                                                       0, & \Theta\neq\pi\mod 2\pi \\
                                                       \frac{1}{2}|eg|, & \Theta=\pi \mod 2\pi
                                                     \end{array}\right.
\ee
and
\be
\Delta(T\rightarrow0;\,\hat{O}_{{\bf J}^2},\,\hat{O}_{{\bf J}^2})=\left\{\begin{array}{cc}
                                                       0, & \Theta,\neq\pi \mod 2\pi\\
                                                       \frac{1}{2}|eg|[(eg)^2-\frac14]^2, & \Theta=\pi\mod 2\pi
                                                     \end{array}.\right.
\ee

\section{Summary}

It is well known~\cite{Wit,Gro,Yam,Cor,Dun} that the vacuum and
thermal fluctuations of the quantized Dirac electron field in the
background of a pointlike magnetic monopole result in the monopole
becoming a dyon with electric charge $eO_I$ depending on the CP
violating vacuum angle, see Eqs.(66) and (95). In the
present study we find out that, in addition to charge, also other
quantum numbers are induced in the monopole background. These
comprise squares of orbital angular momentum, spin, and total
angular momentum, and we show that they are related to
charge, see Eqs.(72)-(74) and Eqs.(96)-(98). The density of induced
quantum numbers is considerable around a monopole in the region of
order of the Compton size of the electron, decreasing exponentially at
larger distances (as $r^{-5/2}e^{-2Mr}$ at $r\rightarrow\infty$), see Eq.(63).

The conserved
observables are charge and squared total angular momentum; note that
the latter vanishes in the case of the minimal monopole strength,
$|eg|=\frac12$. We analyze thermal correlations between conserved and
nonconserved observables and thermal quadratic fluctuations of
conserved observables, and find out that these quantities at nonzero
temperature are given by the ideal gas expressions, see
Eqs.(100)-(106), and, thus, are $\Theta$-independent and
proportional to the powers of spatial volume. The interaction with
the monopole background reveals itself at zero temperature, yielding
a $\Theta$-dependence of a specific type, which is due to a
possibility of appearance of a bound state with zero energy in
the one-particle electron spectrum, see Eq.(113). This fact has
immediate consequences when we turn to a question: whether the
values of charge and squared total angular momentum at zero temperature
are observed in a single quantum measurement, or whether they are to
be regarded as expected averages of many such measurements.

We recall that CP invariance is violated, unless \be \Theta=n\pi.\ee
Induced vacuum quantum numbers, as functions of the vacuum angle,
are discontinuous at points $\Theta=\pi\mod 2\pi$ (i.e. when the
bound state with zero energy appears in the one-particle electron
spectrum), otherwise they are continuous, vanishing at points
$\Theta=0\mod 2\pi$. Since the electric charge of a dyon in the case
of CP conservation can be either integer or half-integer in units of
$e$, this dictates that the induced charge and all other quantum
numbers in the case of Eq.(116) have to take the same, i.e. equal to
zero, values. In other reasoning, it is sufficient to choose range
$|\Theta|\leq \pi$, where end points $\Theta=\pi$ and $\Theta=-\pi$
have to be equivalent, and the equivalence obliges to choose the
mean between the right and left limiting values, i.e. zero value for
the induced quantum number. Also, if we start from nonzero
temperatures, when the induced quantum numbers are continuous in
$\Theta$ everywhere, see Eq.(93) or (94), and tend temperature to
zero, then we get the induced vacuum quantum numbers which are
vanishing at $\Theta=\pi\mod 2\pi$. However, as it follows from the
expressions for quadratic fluctuations at zero temperature,
Eqs.(114) and (115), charge and squared total angular momentum are
sharp observables (quantum-mechanical eigenvalues), unless
$\Theta=\pi\mod 2\pi$. Thus, CP conserving values of the vacuum
angle, Eq.(116), differ significantly: in contrast to the case of
$\Theta=0\mod 2\pi$, charge and squared total angular momentum in
the case of $\Theta=\pi\mod 2\pi$ are expected average values, not
eigenvalues.

\renewcommand{\thesection}{Acknowledgements}
\renewcommand{\theequation}{\thesection.\arabic{equation}}
\setcounter{section}{1} \setcounter{equation}{0}

\section*{Acknowledgements}

This research was supported in part by the Swiss National Science
Foundation under the SCOPES project No.~IB7320-110848. Yu.A.S.
acknowledges the support of the State Foundation for Basic Research
of Ukraine (grant No.~2.7/00152) and INTAS (grant
No.~05-1000008-7865). N.D.V. acknowledges the INTAS support through
the PhD Fellowship Grant for Young Scientists (No.~05-109-5333).

\end{document}